\documentclass[twocolumn,aps,aps,amsfonts,floats]{revtex4}
\usepackage{graphicx}
\usepackage{amsmath}
\usepackage{float}

\begin{document}

\title{The growth of structure in interacting dark energy models}

\author{Gabriela Caldera-Cabral}
\email{Gaby.Calderacabral@port.ac.uk}
\author{Roy Maartens}
\email{Roy.Maartens@port.ac.uk}
\affiliation{Institute of Cosmology \& Gravitation, University of
Portsmouth, Portsmouth~PO1 3FX, UK }
\author{Bjoern Malte Schaefer}
\email{spirou@ita.uni-heidelberg.de}
\affiliation{Astronomisches Recheninstitut, Zentrum f{\"u}r Astronomie,
Universit{\"a}t Heidelberg, M{\"o}nchhofstra{\ss}e 12, 69120
Heidelberg, Germany}

\date{\today}

\begin{abstract}
If dark energy interacts with dark matter, there is a change in the background evolution of the universe, since the dark matter density no longer evolves as $a^{-3}$. In addition, the non-gravitational interaction affects the growth of structure. In principle, these changes allow us to detect and constrain an interaction in the dark sector. Here we investigate the
growth factor and the weak lensing signal for a new class of
interacting dark energy models. In these models, the interaction generalises the simple cases where one dark fluid decays into the other.
In order to calculate the effect on structure formation, we perform a careful analysis of the perturbed interaction and its effect on peculiar velocities.
Assuming a normalization to today's values of dark matter density and overdensity, the signal of the interaction
is an enhancement (suppression) of both the growth
factor and the lensing power, when the energy
transfer in the background is from dark matter to dark energy (dark energy to dark matter).
\end{abstract}

\maketitle 
\section{Introduction}

Advances in cosmological observations have led to strong evidence
for non-baryonic cold dark matter and for a late-time acceleration
of the universe, possibly driven by a dark energy field (see,
e.g.~\cite{Dunkley:2008ie,Tegmark:2006az,Percival:2007yw}). Dark
energy and dark matter are the dominant sources in the `standard'
model for the evolution of the universe. Both are considered
essential missing pieces in the cosmic puzzle -- and both are
currently only detected via their gravitational effects. There
could therefore be an interaction between them without violating
current observational constraints. Furthermore, such an
interaction could alleviate the `coincidence' problem (why are the
energy densities in the two components of the same order of
magnitude today?). And interacting dark energy, by exerting a
non-gravitational `drag' on dark matter, introduces new features
to structure formation, including possibly a new
bias~\cite{Amendola:2001rc} and a violation by dark matter of the
weak equivalence principle~\cite{Bertolami:2007zm,KoyamaMaartensSong-inprep}.

In order to pursue the possibility of interacting dark energy, we
need to compute the effect of the interaction on the background
expansion and on structure formation, and then to confront the
results with data. The main problem is that there is no
fundamental theory to guide us as to the form of an interaction.
This problem is in fact subsidiary to a bigger problem -- that
there is currently no fundamental theory to underpin {\em any}
model of dark energy, including non-interacting dark energy. The situation is
somewhat similar to that in reheating of the universe after
inflation: there is no fundamental theory for the inflaton field,
and no fundamental theory to guide us as to the form of the
interactions between the inflaton and other fields during
reheating. In this situation, we are forced to adopt
phenomenological models to explore and narrow down the space of
possibilities.

Various interacting dark energy models have been put forward (see,
e.g.~\cite{Wetterich:1994bg,Amendola:1999qq,Billyard:2000bh,
Zimdahl:2001ar,Farrar:2003uw,Chimento:2003iea,Olivares:2005tb,Koivisto:2005nr,
Sadjadi:2006qp,Guo:2007zk,Boehmer:2008av,He:2008tn,Quartin:2008px,Pereira:2008at,
Quercellini:2008vh,Valiviita:2008iv,He:2008si,Bean:2008ac,Chongchitnan:2008ry,
Corasaniti:2008kx,Gavela:2009cy,Jackson:2009mz}). All of these models are phenomenological. Some of them are constructed specifically for mathematical simplicity -- for example, models in which the energy exchange rate is proportional to the Hubble rate. We consider models which are similar to simple models of reheating and of curvaton decay, i.e., where the energy exchange is in the form of a decay of one species into another: the energy exchange is linear in the energy density of the decaying species, and the decay rate is constant. Such models were introduced in~\cite{Boehmer:2008av,Valiviita:2008iv}, where dark matter decays to dark energy. Here we follow~\cite{CalderaCabral:2008bx} and generalize the energy exchange to a linear
combination of dark sector energy densities, including as special cases the decay of dark matter to dark energy, and the decay of dark energy to dark matter.

We analyze the growth rate of structure and the weak lensing
convergence, both of which are sensitive to
interacting dark energy. In a previous work~\cite{Schaefer:2008ku}, we considered the case of dark matter decaying to dark energy. Here we generalize to include the case of dark energy decaying to dark matter. In~\cite{Schaefer:2008ku} we did not analyze the peculiar velocity of dark matter, but rather deduced the density perturbation evolution using qualitative arguments. Here we perform a a careful analysis of the peculiar velocity, which confirms the qualitative arguments of~\cite{Schaefer:2008ku} -- but also allows us to deal with the new case of dark energy decay, where the peculiar velocity analysis is essential for deriving the correct density perturbation evolution.

Following the typical approach in
non-interacting dark energy models, we use the parametrization of the equation of state for the dark energy~\cite{Chevallier:2000qy,Linder:2003dr},
\begin{equation}
w(a)=w_0+w_a(1-a)\,.\label{wa}
\end{equation}

In the background, a general coupling can be described by the
continuity equations of cold dark matter (c) and dark energy (x),
\begin{eqnarray}
\dot{\bar{\rho}}_c+3H\bar{\rho}_c&=& \bar{Q}_c \,,
\label{rhocb}\\
\dot{\bar{\rho}}_x+3H(1+w)\bar{\rho}_x&=& \bar{Q}_x= -\bar{Q}_c
\,, \label{rhoxb}
\end{eqnarray}
where $w=\bar{P}_x/\bar{\rho}_x$, the background metric is $
ds^2=-dt^2+a^2d\vec x\,^2$, and $\bar Q_c~(\bar Q_x)$ is the rate
of energy density transfer to dark matter (dark energy). Therefore
$\bar Q_c<0~(>0)$ implies that the direction of energy transfer is
dark matter $\rightarrow$ dark energy (dark energy $\rightarrow$
dark matter).

In order to avoid stringent ``fifth force" constraints, we assume
that baryons (b) and photons ($\gamma$) are not coupled to dark
energy: $\bar{Q}_b= 0 = \bar Q_\gamma$. We also neglect radiation since we are focusing on structure
formation in the late universe. The energy balance equation for fluid $A$ is
\begin{equation}\label{rhoA}
\dot{\bar{\rho}}_A+3H(1+w_A)\bar{\rho}_A= \bar{Q}_A\,,
\end{equation}
with $ w_c=w_b=0,w_x=w$ and $\bar{Q}_c=-\bar{Q}_x\neq 0=
\bar{Q}_b$. The Friedmann equation is
\begin{equation}\label{f}
H^2=\frac{8\pi G}{3}(\bar{\rho}_c+\bar{\rho}_x+\bar{\rho}_b).
\end{equation}

\subsection*{Background dynamics}

Once a form of $\bar Q_c$ is given, the background dynamics are
fully determined by Eqs.~(\ref{rhoA}) and (\ref{f}), with $w$
given by Eq.~(\ref{wa}). Here we use the new
model~\cite{CalderaCabral:2008bx}
\begin{equation} \label{Q}
\bar Q_c=-\left(\Gamma_c\bar{\rho}_c+\Gamma_x\bar{\rho}_x \right)
\,,
\end{equation}
where $\Gamma_A$ are constant energy density transfer rates. The special cases of Eq.~(\ref{Q}) are: decay of dark matter to dark energy, i.e., $\Gamma_c>0 = \Gamma_x$, and decay of dark energy to dark matter, i.e., $\Gamma_x<0 = \Gamma_c$.
Models of the pure-decay type have been used in reheating after
inflation~\cite{Turner:1983he}, and to describe the decay of dark
matter into radiation~\cite{Cen:2000xv} or of a curvaton field into
radiation~\cite{Malik:2002jb}.

A complete dynamical systems analysis for Eq.~(\ref{Q}) is given
in~\cite{CalderaCabral:2008bx} for the case $w=\,$const, i.e.,
$w_a=0$ in Eq.~(\ref{wa}). In this paper we will consider variable
$w$, and the particular cases when one of the transfer rates
$\Gamma_A$ is zero. The special case previously considered
in~\cite{Valiviita:2008iv} corresponds to $\Gamma_x=0$, with
$w_a=0$. As shown in~\cite{Valiviita:2008iv}, when $\Gamma_c>0$ and
$w_a=0$, the dark energy density becomes negative at early times.
The source of this problem is the rigidity of the assumption that $w$ is constant
-- the problem does not arise for quintessence models. It can also
be avoided with variable $w$ for suitable choices of $w_0, w_a$ in
Eq.~(\ref{wa}), as shown in~\cite{ValiviitaMaartensMajerotto-in-prep}.

The case $\Gamma_c<0$ (with $w_a=0$) avoids negative dark energy
density, but there is no attractor solution; by contrast, there is
a late-time attractor when
$\Gamma_c>0$~\cite{CalderaCabral:2008bx}.
Equation~(\ref{rhocb}) has an exact solution for $\Gamma_x=0$,
\begin{equation}
\bar\rho_c=\bar\rho_{c0}a^{-3}\exp[-\Gamma_c(t-t_0)],
\end{equation}
which shows that the dark matter density is always positive,
regardless of the sign of $\Gamma_c$. For the special case
$\Gamma_c=0 = w_a$, the dark energy density is always positive, as can
be seen from the exact solution of Eq.~(\ref{rhoxb}):
\begin{equation}
\bar\rho_x=\bar\rho_{x0}a^{-3(1+w)}\exp[\Gamma_x(t-t_0)].
\end{equation}

\section{Density and velocity perturbation equations}

Since we are interested in the late universe we can neglect
anisotropic stress, and scalar perturbations of the flat metric
are given, in Newtonian gauge, by
\begin{equation}
ds^2=-(1+2\phi)dt^2+a^2(1-2\phi)d\vec{x}^2.
\end{equation}
The $A$-fluid four-velocity is
\begin{equation}
u^{\mu}_A=\left(1-\phi,\partial^iv_A\right).
\end{equation}
Choosing the energy frame for the total four-velocity $u^{\mu}$,
the total velocity potential $v$ is defined by
\begin{equation}
v \sum\left(\bar{\rho}_A +
\bar{P}_A\right)=\sum\left(\bar{\rho}_A+ \bar{P}_A\right)v_A\,.
\end{equation}
The covariant form of energy-momentum transfer is
\begin{equation}
\nabla_{\nu}T^{\mu\nu}_A=Q^{\mu}_A\,,~~ Q^{\mu}_c=-Q^{\mu}_x\neq
0=Q^\mu_b\,.
\end{equation}
The energy-momentum transfer four-vector can be split relative to
the total four-velocity as~\cite{Valiviita:2008iv}
\begin{equation}
Q^{\mu}_A=Q_Au^{\mu}+F^{\mu}_A,~ Q_A=\bar Q_A+\delta Q_A,~
u_{\mu}F^{\mu}_A=0,
\end{equation}
where $F^{\mu}_A$ is the momentum density transfer rate, relative
to $u^{\mu}$. Then it follows that $F^{\mu}_A=\left(0,\partial^i
f_A\right)$, where $f_A$ is a momentum transfer potential, and
\begin{eqnarray}
Q^A_0&=&-\left[\bar{Q}_A\left(1+\phi\right)+\delta Q_A\right],
\label{q_mom_1}\\
Q^A_i&=&a^2\partial_i\left(f_A+\bar{Q}_Av\right). \label{q_mom_2}
\end{eqnarray}

The general evolution equations for the dimensionless density
perturbations $\delta_A=\delta\rho_A/\rho_A$ and the velocity
perturbations $v_A$ are~\cite{Valiviita:2008iv}:
\begin{eqnarray}
&& \dot\delta_A+3Hc^2_{sA}\delta_A - (1+w_A)\frac{k^2}{a}v_A
\nonumber\\ &&{} -3H[3H(1+w_A)(c^2_{sA}-w_A)+\dot w_A]v_A
-3(1+w_A)\dot\phi \nonumber \\&&{} =\frac{\delta
Q_A}{\bar{\rho}_A}+\frac{\bar{Q}_A} {\bar{\rho}_A}[\phi-\delta_A-
3aH(c^2_{sA}-w_A)v_A], \label{da}\\
&& \dot v_A+H(1-3c^2_{sA})v_A+\frac{c^2_{sA}}{a(1+w_A)}
\delta_A+\frac{\phi}{a} \nonumber\\&&{} =
\frac{1}{(1+w_A)\bar{\rho}_A}
\Big\{\bar{Q}_A\left[v-(1+c^2_{sA})v_A\right]+ f_A\Big\}.
\label{va}
\end{eqnarray}
The relativistic Poisson equation is:
\begin{equation}
\frac{k^2 \phi}{a^2}=-3H\dot\phi-3H^2\phi-4\pi G\left(
\bar{\rho}_c\delta_c + \bar{\rho}_b\delta_b +
\bar{\rho}_x\delta_x\right). \label{poisson_complete}
\end{equation}

Structure formation takes place in the Newtonian regime, on
spatial scales much smaller than the horizon radius, $a/k \ll
H^{-1} $. Then the gravitational potential and its time derivative may
be neglected relative to matter density fluctuations. This allows us to discard the $\phi$ terms in Eq.~(\ref{da}) and the first two terms on the right of Eq.~(\ref{poisson_complete}). Dark energy
fluctuations may be neglected since dark energy has a high sound
speed, and does not cluster on sub-Hubble scales (we take the
sound speed of dark energy to be that of a standard scalar field
model, $ c_{sx}^2=1$). In the Newtonian regime, the evolution equations for
dark matter become:
\begin{eqnarray}
\dot\delta_c-\frac{k^2}{a}v_c &=&\frac{1}{\bar{\rho}_c}
\left(\delta Q_c-\bar{Q}_c\delta_c\right), \label{density_new} \\
\dot v_c+Hv_c+\frac{1}{a}\phi &=&
\frac{1}{\bar{\rho}_c}\left[\bar{Q}_c(v-v_c)+f_c\right],
\label{velocity_new}
\end{eqnarray}
while for baryons:
\begin{eqnarray}
\dot\delta_b-\frac{k^2}{a}v_b &=& 0\,,\label{db} \\
\dot v_b+Hv_b+\frac{1}{a}\phi &=& 0\,. \label{vb}
\end{eqnarray}
The Poisson equation in the Newtonian limit and neglecting dark energy clustering is
\begin{equation}
\frac{k^2}{a^2}\phi=-4\pi G\left( \bar{\rho}_c\delta_c +
\bar{\rho}_b\delta_b \right). \label{poisson_new}
\end{equation}

\subsection*{Dark sector energy-momentum transfer}

In order to analyze structure formation in interacting dark energy
models, we need to specify a covariant form of the transfer
four-vectors, $ Q^\mu_c=- Q^\mu_x$, which recovers the energy
transfer model of Eq.~(\ref{Q}) in the background.

Firstly, following and generalizing~\cite{Valiviita:2008iv}, we
can promote Eq.~(\ref{Q}) to the perturbed universe directly via
\begin{equation} \label{qc}
Q_c:=\bar{Q}_c+\delta Q_c =-\Gamma_c \bar{\rho}_c(1+ \delta_c)
-\Gamma_x \bar{\rho}_x(1+ \delta_x).
\end{equation}

However, the background model provides no guide for specifying the
momentum transfer (which vanishes in the background). It is crucial to provide a covariant prescription that fixes the momentum transfer, a point that is not always clearly recognised in the literature.
Following~\cite{Valiviita:2008iv}, we specify the momentum
transfer by the covariant physical requirement that there is no
momentum transfer in the dark matter frame, i.e.,
\begin{equation}
Q_c^{\mu}=Q_c u^{\mu}_c \,. \label{form}
\end{equation}
Then Eqs.~(\ref{qc}) and (\ref{form}) completely describe the
interaction.

Regardless of the specific form of $Q_c$, it follows from
Eq.~(\ref{form}), using Eqs.~(\ref{q_mom_1}) and (\ref{q_mom_2}),
that
\begin{equation}
f_c=\bar{Q}_c\left(v_c-v\right)\,,
\end{equation}
so that Eq.~(\ref{velocity_new}) becomes
\begin{equation}
\dot v_c+Hv_c+\frac{\phi}{a}=0 \,. \label{velocity_new_2}
\end{equation}
This means that the dark matter velocity is the same as the baryon
velocity, and is not directly affected by the interaction with
dark energy. In particular, the form of energy-momentum transfer
in Eq.~(\ref{form}), for any $Q_c$, ensures that there is no
violation of the weak equivalence principle for dark matter, as
shown in~\cite{KoyamaMaartensSong-inprep}. This will not be true
for interaction models where $Q_c^\mu$ is not parallel to
$u^\mu_c$.

\section{Growth factor and weak lensing: the case $\Gamma_x= 0$}

For this model,
\begin{equation}
Q_c^\mu=-\Gamma_c\bar{\rho}_c(1+\delta_c)\,u^\mu_c\,.
\end{equation}
It follows from Eqs.~(\ref{rhocb}) and (\ref{rhoxb}) that in the
background, $\Gamma_c>0$ corresponds to the decay of dark matter
into dark energy, while $\Gamma_c<0$ describes energy transfer
from dark energy to dark matter.

The equations for density and velocity perturbations and the
Poisson equation are
\begin{eqnarray}
\dot\delta_c-\frac{k^2v_c}{a}&=&0\,,\\
\dot v_c+Hv_c+\frac{\phi}{a}&=&0\,,\\
k^2\phi&=&-4\pi Ga^2(\bar\rho_c\delta_c + \bar\rho_b\delta_b).
\end{eqnarray}
From these equations, the second-order evolution equation for
$\delta_c$ is
\begin{equation}
\ddot \delta_c+2H\dot \delta_c-4\pi G (\bar\rho_c\delta_c +
\bar\rho_b\delta_b)=0 \,. \label{evolutionc}
\end{equation}

These equations have precisely the same form as those for the
non-interacting case. However, the {\em solutions} $ \delta_c$ are different because
the background terms $H$ and $\bar\rho_c$ evolve differently. The
interaction will produce a signal in the growth of structure, as
measured by the growth factor and weak lensing
measurements. Equation~(\ref{evolutionc}) confirms the qualitative arguments used in~\cite{Schaefer:2008ku}, where the velocity perturbations were not analyzed. In this section, we extend the analysis of~\cite{Schaefer:2008ku} to include the case $\Gamma_c<0$.

For our computations, the initial conditions are set at the
present epoch $a_0=1$, and we integrate backwards to $a=10^{-2}$. We normalize the dark matter background densities and density perturbations in the interacting and non-interacting cases to today's values, $\Omega_{c0}$ and $\delta_{c0}$.
We take $\Gamma_c=\pm 0.3H_0$, and use an equation of state~(\ref{wa}) with
$w_0=-0.99$ and $w_a=0.8$. These values are
known~\cite{ValiviitaMaartensMajerotto-in-prep} to avoid
$\bar\rho_x<0$ in the past when $\Gamma_c>0$. For simplicity, and
since we are only illustrating the effects of interacting dark
energy, rather than making accurate parameter estimations, we will
neglect the baryons in our computations.

\begin{figure*}[ht!]
\centering
\includegraphics[width=0.50\textwidth]{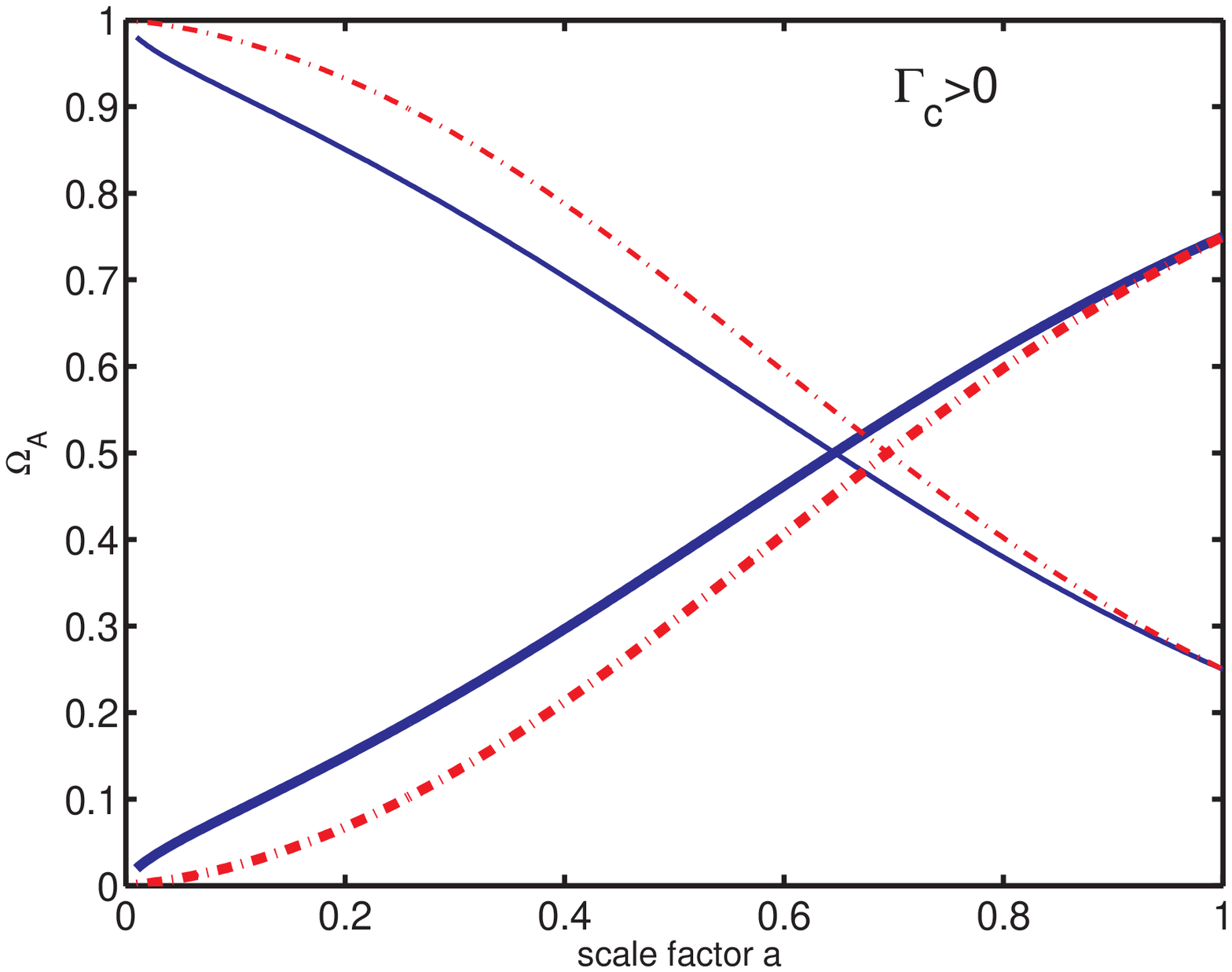}\hfill
\includegraphics[width=0.50\textwidth]{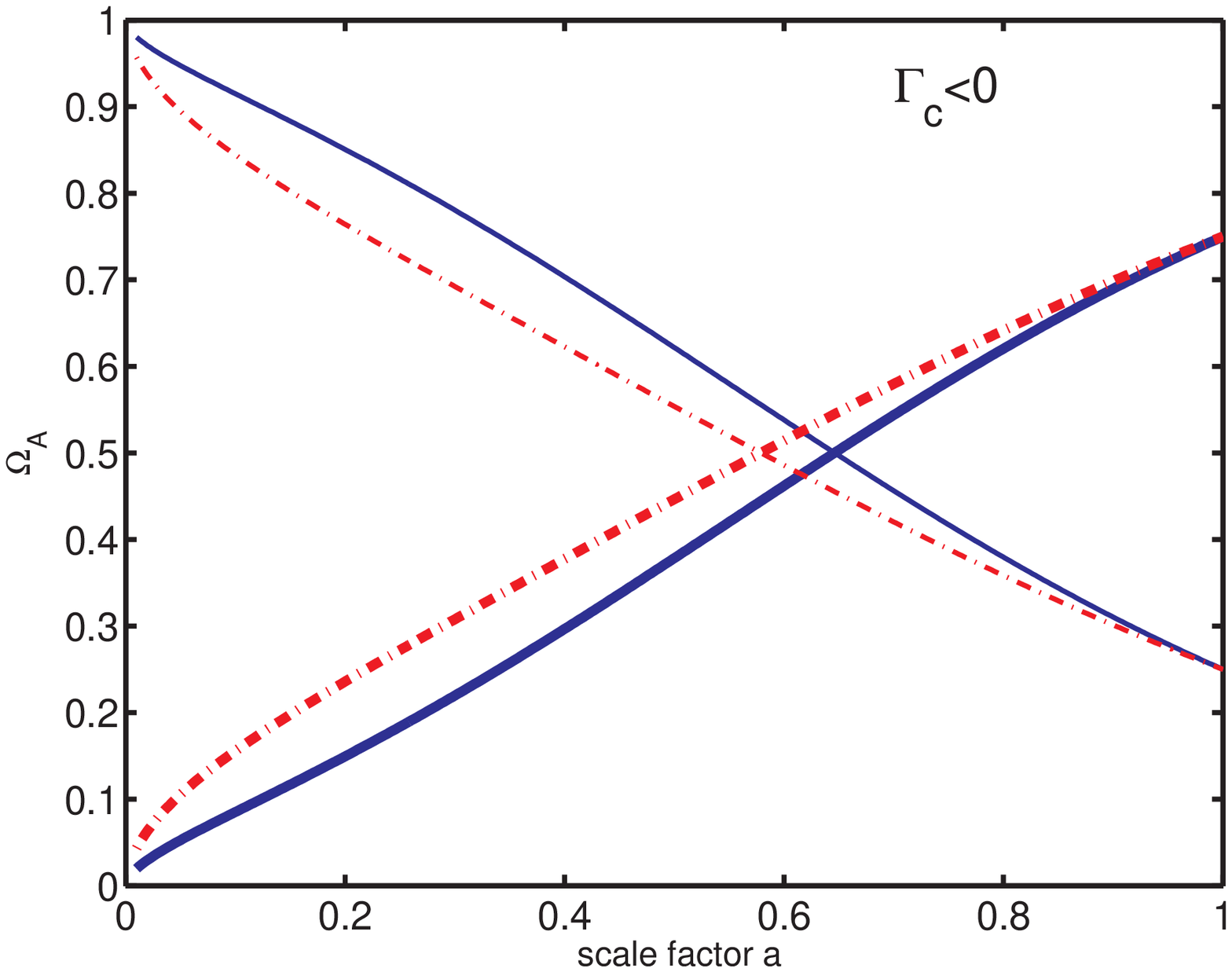}
\caption{The densities of dark matter, $\Omega_c$ (thin lines),
and dark energy, $\Omega_x$ (thick lines), in the interacting
models (dashed-dotted lines), with $\Gamma_c=\pm 0.3H_0$, and
non-interacting models (solid lines), normalized to today's values.} \label{densityc}
\end{figure*}

In Fig.~\ref{densityc} we plot the energy densities $\Omega_A=8\pi
G \bar\rho_A/3H^2 $. Models with dark matter decay $(\Gamma_c>0)$
have higher dark matter density in the past relative to the
non-interacting models (recall that we normalize the densities in the two
models to the same values at $a_0=1$). By contrast, models with
energy transfer from dark energy to dark matter $(\Gamma_c<0)$
exhibit lower dark matter density in the past.

\begin{figure*}[ht!]
\centering
\includegraphics[width=0.50\textwidth]{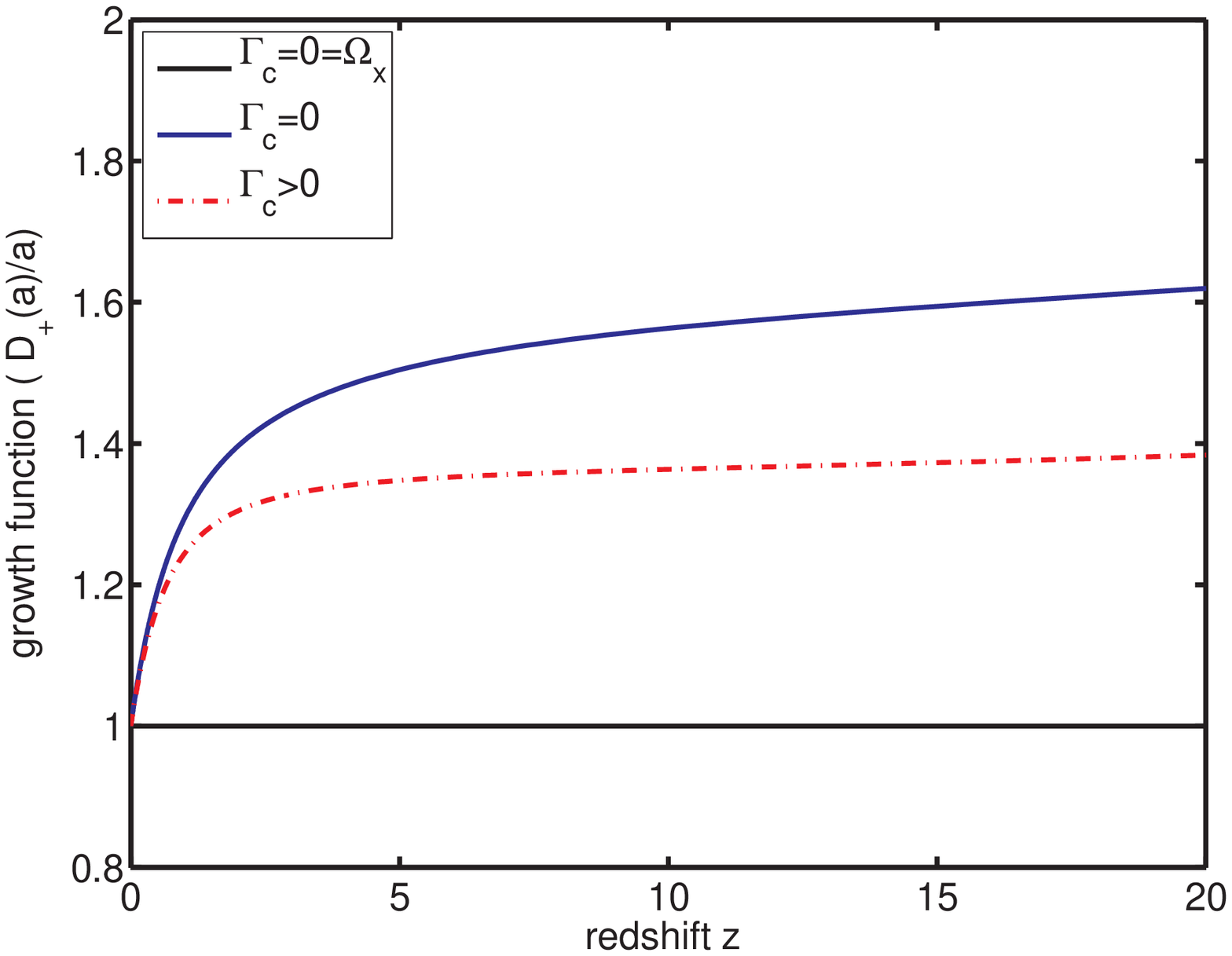}\hfill
\includegraphics[width=0.50\textwidth]{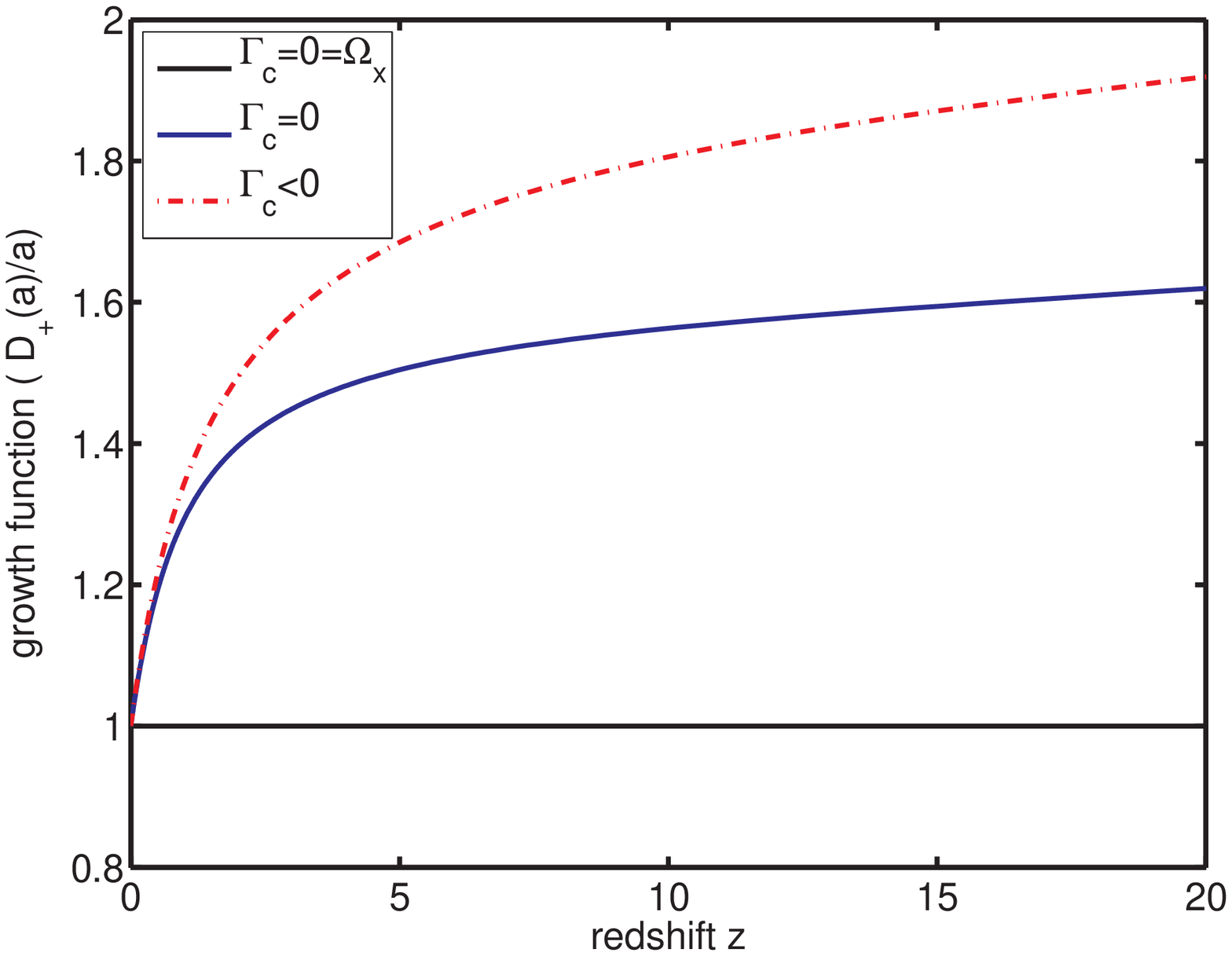}
\caption{Linear growth function $D_+=\delta_c/\delta_{c0}$,  normalized to today's value, relative to its value in a pure-matter model ($D_+=a$). The
interacting models (dashed-dotted lines), with $\Gamma_c=\pm
0.3H_0$, are shown in comparison to non-interacting models (solid
lines).} \label{growthc}
\end{figure*}

Figure~\ref{growthc} shows the growth function $D_+
=\delta_c/\delta_{c0}$ in the interacting models compared to the
non-interacting case, normalized to today's value. The quantity $D_+/a$  reflects the time evolution of the gravitational potential $\phi$. The model with
decaying dark matter, $\Gamma_c>0$, shows an enhancement of
structure growth relative to the non-interacting model, since a lower initial value at early times leads to the
same final value today, $\delta_{c0}$, as the non-interacting model. By contrast,
models with $\Gamma_c<0$ show a suppression of structure growth.
These features are consistent with the background evolution shown
in Fig.~\ref{densityc} -- recall that the difference in the growth
function is determined entirely by the different background
evolution, since Eq.~(\ref{evolutionc}) has the same form as the
non-interacting version. When $\Gamma_c>0$, there is more dark
matter in the past, and this leads to an enhancement in the growth
of structure. The reverse holds for $\Gamma_c<0$. Note that the enhancement / suppression is specific to the circumstance that the interacting and non-interacting models are normalized to have the same parameters today, i.e., $\Omega_{c0}$ and $\delta_{c0}$.

The weak lensing convergence spectrum is given
by~\cite{Schaefer:2008ku}
\begin{equation}
C_k(\ell)=\frac{9}{4c^4}\int^{\chi}_0 d{\chi}\,G^2 (aH)^4
\Omega_c^2 D_+^2P(k=\ell/\chi), \label{lensing}
\end{equation}
where $ \chi(a)=\int^1_a\,{da}/{a^2H(a)}$ is the comoving
distance, $ P(k)$ is the dark matter power spectrum, and
\begin{equation}
G(\chi)=\int^{\chi}_0
d\chi'\,p(z)\frac{dz}{d\chi'}\frac{\chi'-\chi}{\chi'}\,,
\end{equation}
with $p(z)$ giving the redshift distribution of lensing galaxies.

\begin{figure*}[ht!]
\centering
\includegraphics[width=0.50\textwidth]{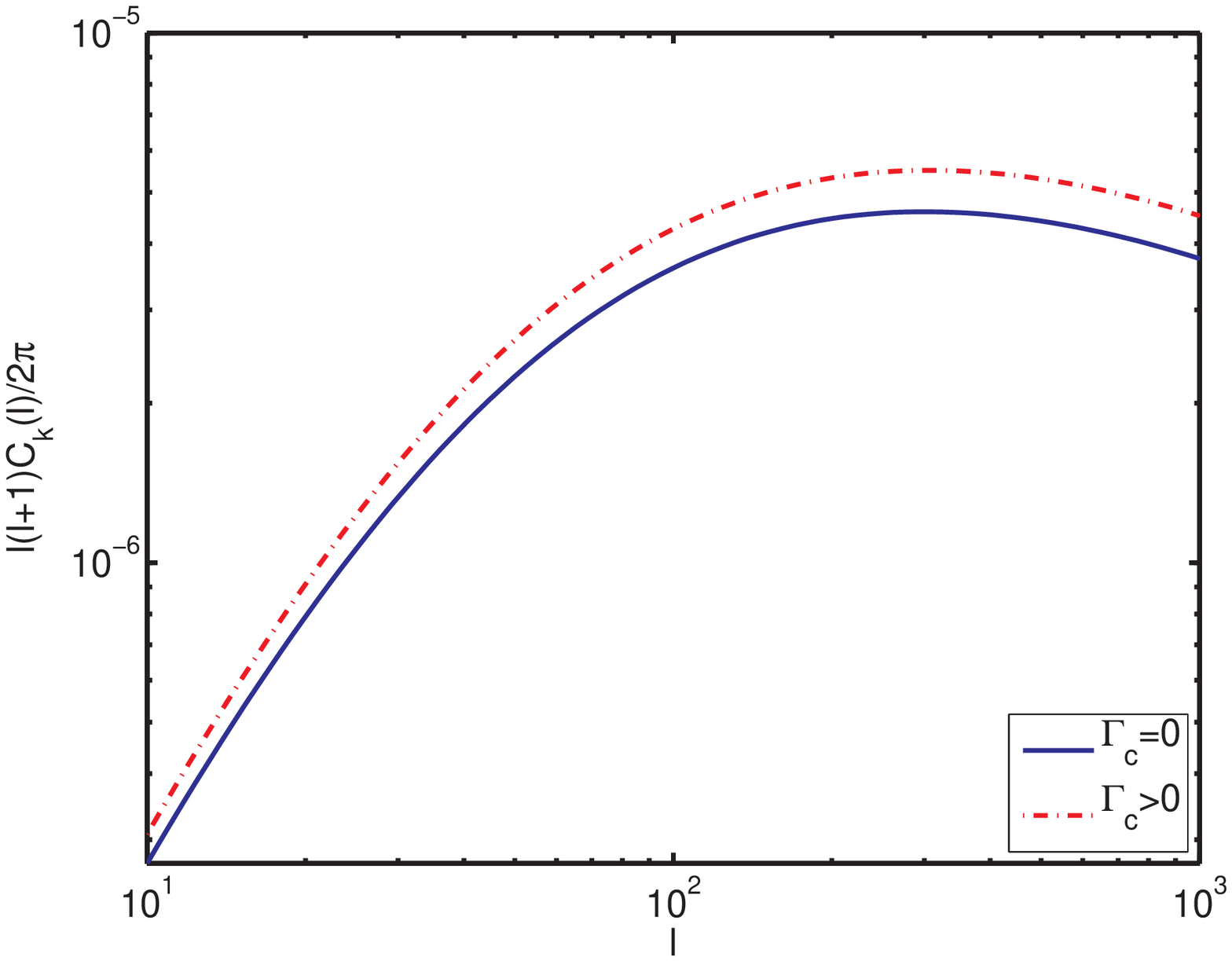}\hfill
\includegraphics[width=0.50\textwidth]{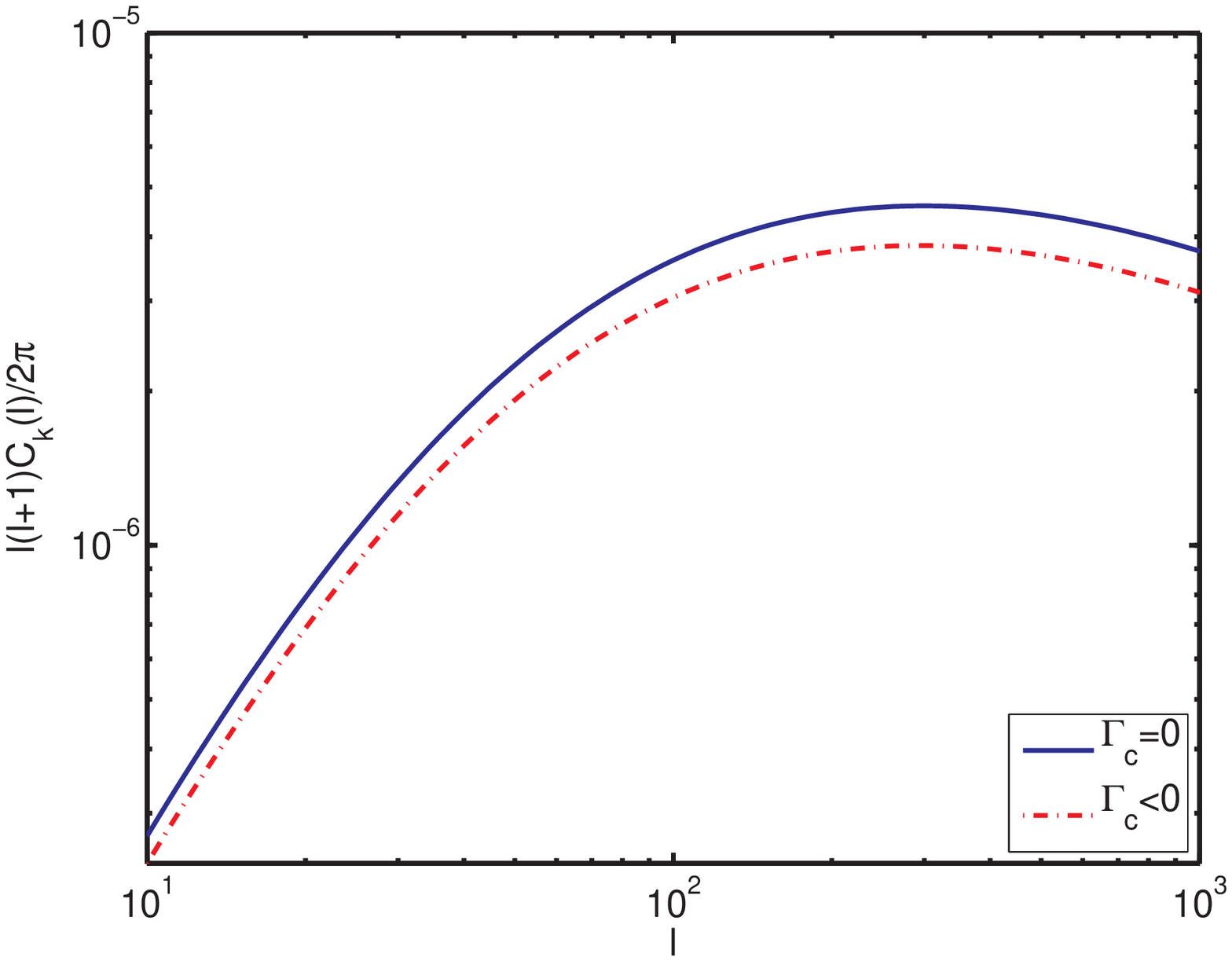}
\caption{Weak lensing convergence power spectra $\ell(\ell+1) C_k(\ell) /2\pi$ in the
interacting models (dashed-dotted lines), with $\Gamma_c=\pm
0.3H_0$, in comparison to non-interacting models (solid lines).}
\label{lensingc}
\end{figure*}

Figure \ref{lensingc} illustrates the impact of interacting dark
energy on the weak lensing convergence power spectra. The lensing power
in models with decaying dark matter ($\Gamma_c>0$) is higher since
the dark matter density in the past was higher (see
Fig.~\ref{densityc}), compared to non-interacting models -- assuming that the models are normalized to have the same values today of $\Omega_{c0}$. The
higher matter density leads to stronger gravitational potential
and hence a stronger light deflection. When $\Gamma_c<0$, the
reverse situation applies.

Our results in the case $\Gamma_c>0$ are consistent with those of~\cite{Schaefer:2008ku}, where in addition the weak lensing bispectrum is computed. The case $\Gamma_c<0$ was not considered in~\cite{Schaefer:2008ku}.

\section{Growth factor and weak lensing: the case $\Gamma_c=0$}

Weak lensing has not previously been analyzed in this model, for which
\begin{equation}
Q_c^\mu=-\Gamma_x\bar{\rho}_x(1+\delta_x)\, u^\mu_c\,.
\end{equation}
It follows from Eqs.~(\ref{rhocb}) and (\ref{rhoxb}) that in the
background, $\Gamma_x<0$ corresponds to the decay of dark energy
into dark matter, while $\Gamma_x>0$ describes energy transfer
from dark matter to dark energy. This model does not present the
problem of negative dark energy density in the past.

The velocity perturbation equation~(\ref{velocity_new_2}) is the
same as in the previous model, but the density perturbation
equation~(\ref{density_new}) has a source term:
\begin{eqnarray}
\dot\delta_c-\frac{k^2v_c}{a}=\Gamma_x\frac{\bar\rho_x}
{\bar\rho_c}\delta_c\,,
\end{eqnarray}
which will generate a linear bias between dark matter and baryons.
This source term leads to a modified evolution equation for
$\delta_c$:
\begin{eqnarray}
&& \ddot \delta_c+2H\left(1-\frac{\Gamma_x}{H}
\frac{\bar\rho_x}{\bar\rho_c}\right) \dot \delta_c -4\pi G\Bigg\{
\bar\rho_c\left[1+ \frac{2}{3a}\frac{\Gamma_x}{H}
\frac{\bar\rho_x}{\bar\rho_c}\times \right. \nonumber\\&&\left.{}
\times \left\{2-3w+\frac{\Gamma_x}{H}\left(1+
\frac{\bar\rho_x}{\bar\rho_c}\right)\right\}\right]\delta_c
+\bar\rho_b\delta_b \Bigg\} =0. \label{evolutionx}
\end{eqnarray}
The evolution equation for dark matter density perturbations is
modified relative to the non-interacting models in 3 ways:
\begin{enumerate}
\item[(1)]
due to the different background evolution (affecting the
terms $\bar\rho_c, \bar\rho_x, H$);
\item[(2)]
due to the modified
Hubble friction term, $H\to H(1-{\Gamma_x \bar\rho_x}
/H\bar\rho_c)$;
\item[(3)]
due to the modified source term, which gives rise to a
modified effective Newton constant $G_{\rm eff}$ for dark matter:
\begin{equation}
{G_{\rm eff} \over G} =
1+\frac{2}{3a}\frac{\Gamma_x}{H}\frac{\bar\rho_x}{\bar\rho_c}
\left\{2-3w+\frac{\Gamma_x}{H}\left(1+\frac{\bar\rho_x}{\bar\rho_c}
\right)\right\},
\end{equation}
so that
\begin{equation}
\ddot \delta_c+2H\!\left(\!1-\frac{\Gamma_x}{H}
\frac{\bar\rho_x}{\bar\rho_c}\! \right)\!\dot \delta_c-4\pi G_{\rm
eff}\bar\rho_c\delta_c -4\pi G \bar\rho_b\delta_b=0\,.
\label{evolutionxred}
\end{equation}
\end{enumerate}

Compared to the previous model, where only modification~(1)
operates, this model shows a more complicated change from the
non-interacting case.

\begin{figure*}[ht!]
\centering
\includegraphics[width=0.50\textwidth]{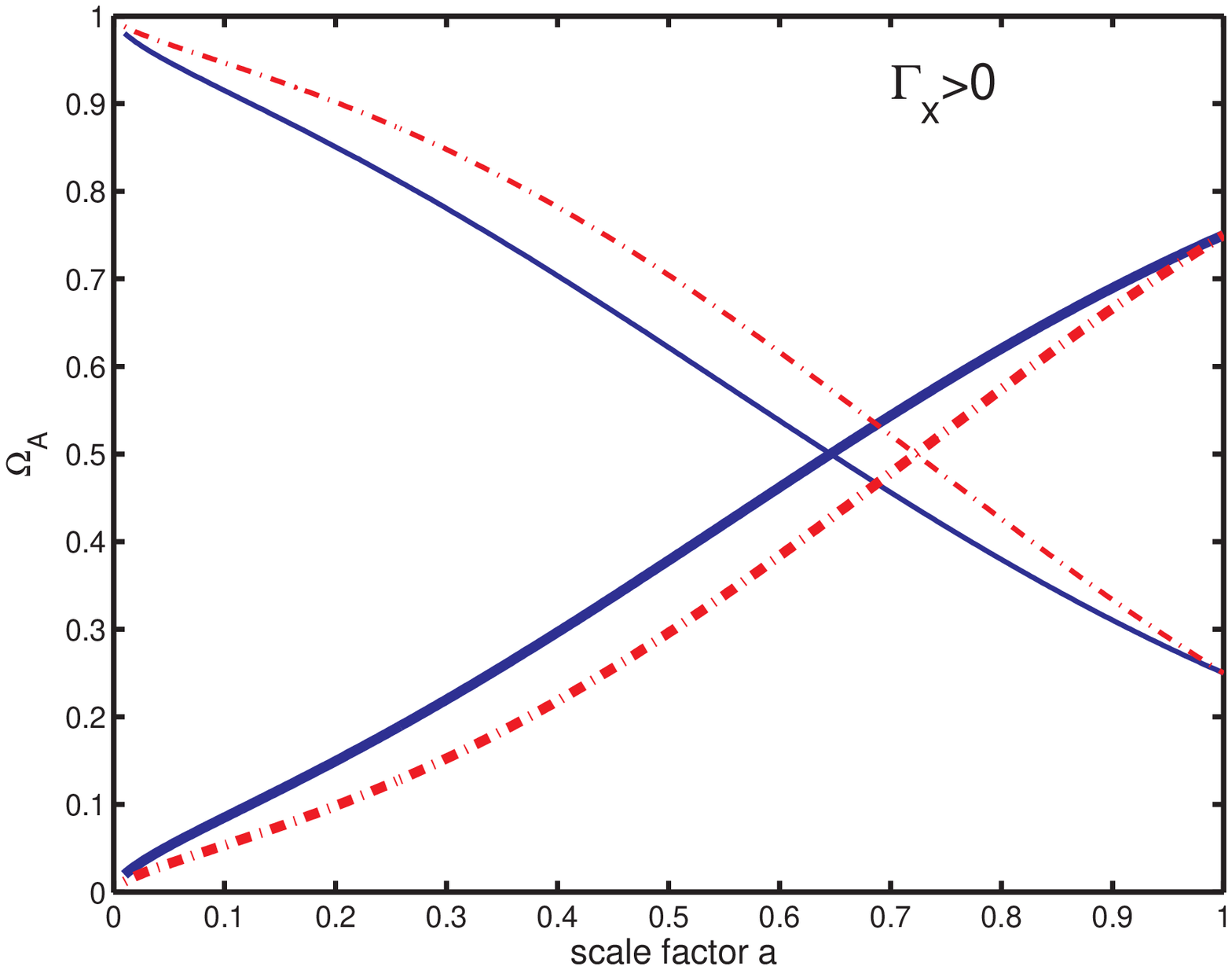}\hfill
\includegraphics[width=0.50\textwidth]{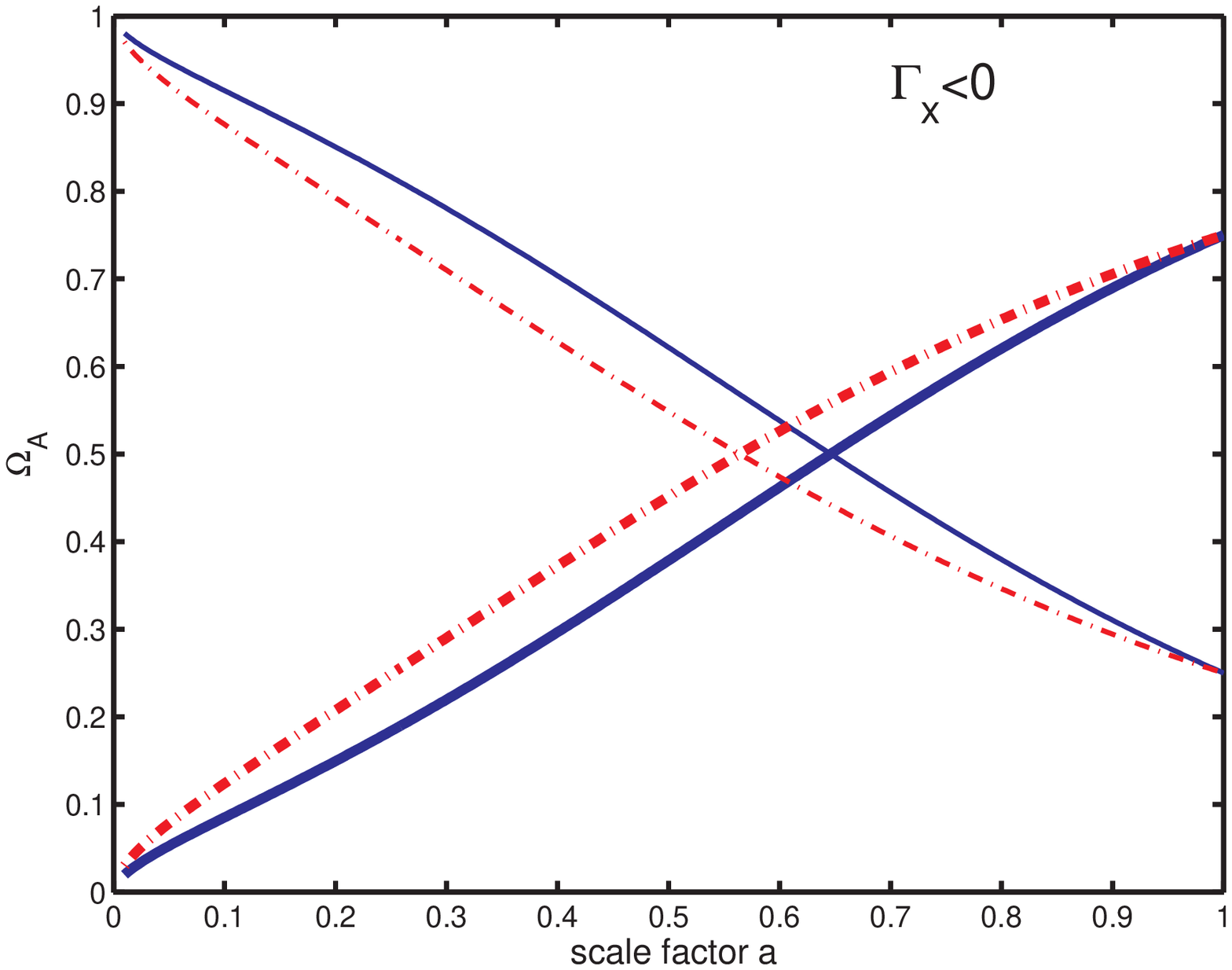}
\caption{The densities of dark matter, $\Omega_c$ (thin lines),
and dark energy, $\Omega_x$ (thick lines), in the interacting
models (dashed-dotted lines), with $\Gamma_x=+ 0.3H_0, -0.2H_0$,
and non-interacting models (solid lines), normalized to today's values.} \label{densityx}
\end{figure*}

Figure~\ref{densityx} shows the energy densities $\Omega_A$. (We
again neglect baryons in the plots.) The model with energy
transfer from dark matter to dark energy $(\Gamma_x>0)$, has
higher dark matter density in the past. The model with dark energy
decay to dark matter $(\Gamma_x<0)$ has lower dark matter density
in the past. (We normalize to today's value of $\Omega_{c0}$.)

\begin{figure*}[ht!]
\centering
\includegraphics[width=0.50\textwidth]{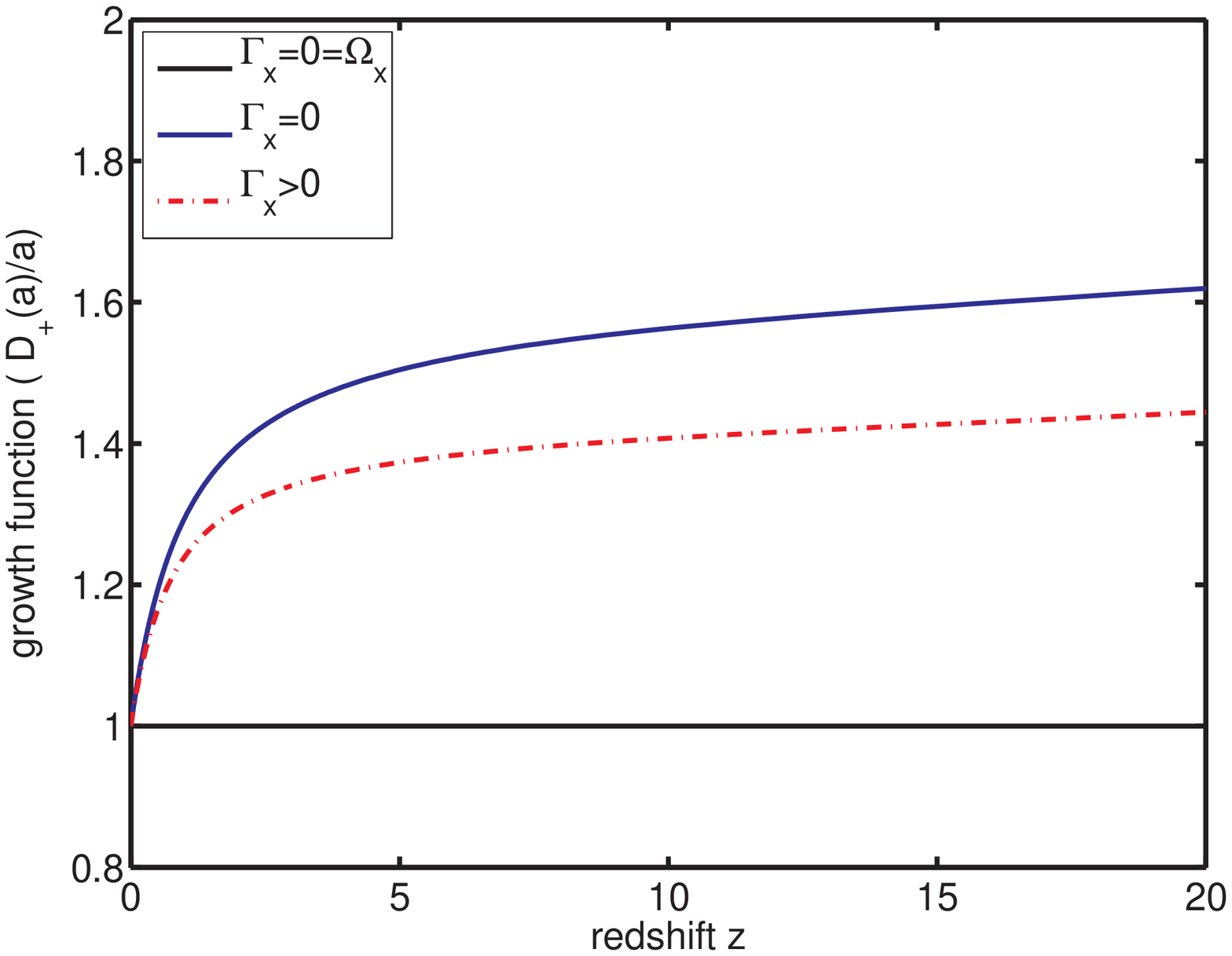}\hfill
\includegraphics[width=0.50\textwidth]{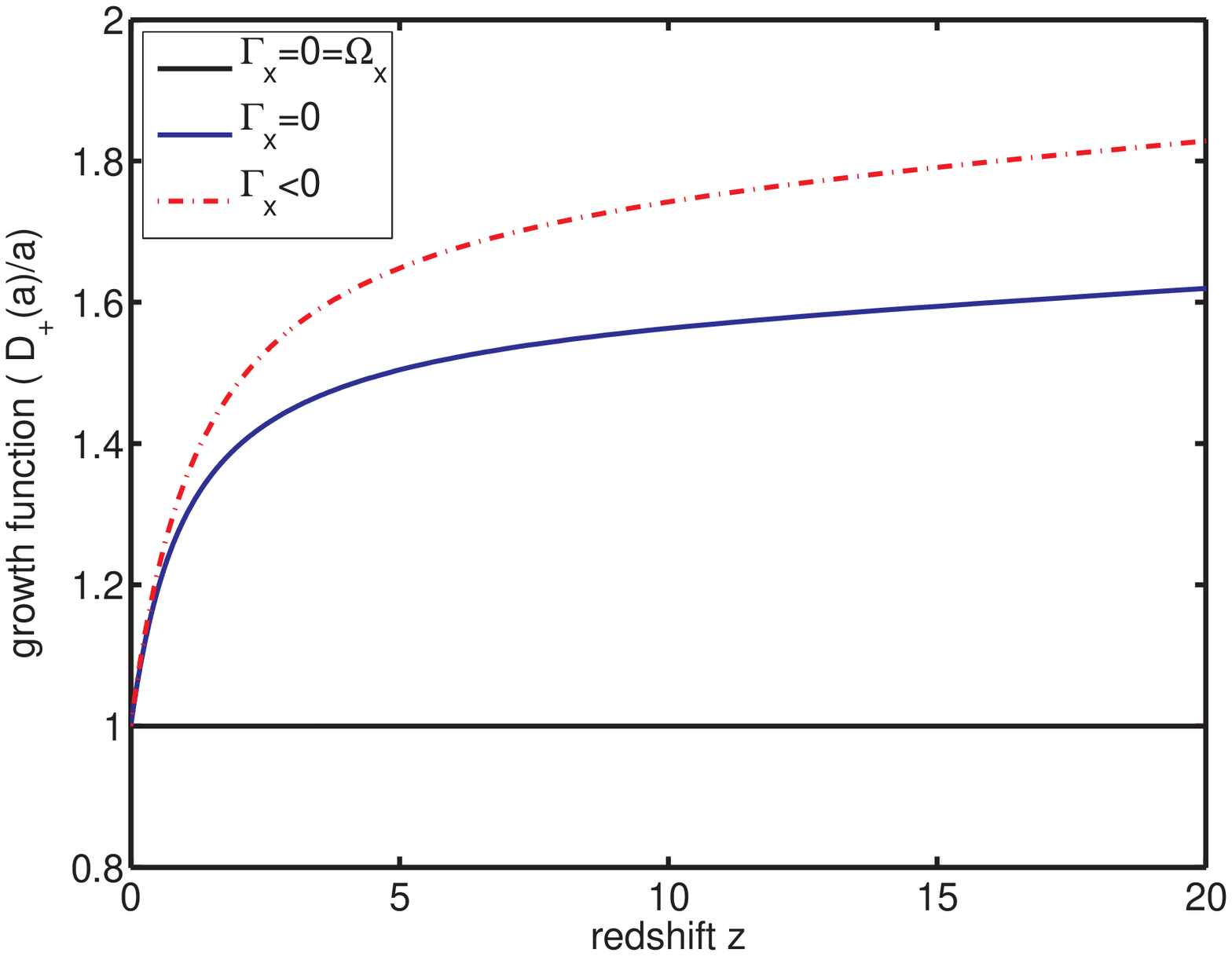}
\caption{Linear growth function $D_+=\delta_c/\delta_{c0}$, normalized to today's value, relative to its value in a pure-matter model ($D_+=a$). The
interacting models (dashed-dotted lines), with $\Gamma_x=+ 0.3H_0,
-0.2H_0 $, are shown in comparison to non-interacting models
(solid lines).} \label{growthx}
\end{figure*}

In Fig.~\ref{growthx} we show the growth function obtained by
solving Eq.~(\ref{evolutionxred}) (neglecting baryons). As in the
previous model, we see that an energy transfer from dark matter to
dark energy, $\Gamma_x>0$, leads to enhancement of growth, since
less initial power is needed to obtain the same final value as the
non-interacting model. The reverse holds for the model with
$\Gamma_x<0$, which suppresses growth for the same final value $\delta_{c0}$. These background effects
are reinforced by the modifications~(2) and (3) to the $\delta_c$ evolution
equation Eq.~(\ref{evolutionxred}):
 \begin{itemize}
\item
When $\Gamma_x>0$, the Hubble friction term is reduced,
and the source term is enhanced, both contributing to an increase
in $\delta_c$ relative to the non-interacting case.
\item
For
$\Gamma_x<0$, the friction term is enhanced and the source term
is reduced, both contributing to a reduction in $\delta_c$.
\end{itemize}

\begin{figure*}[ht!]
\centering
\includegraphics[width=0.50\textwidth]{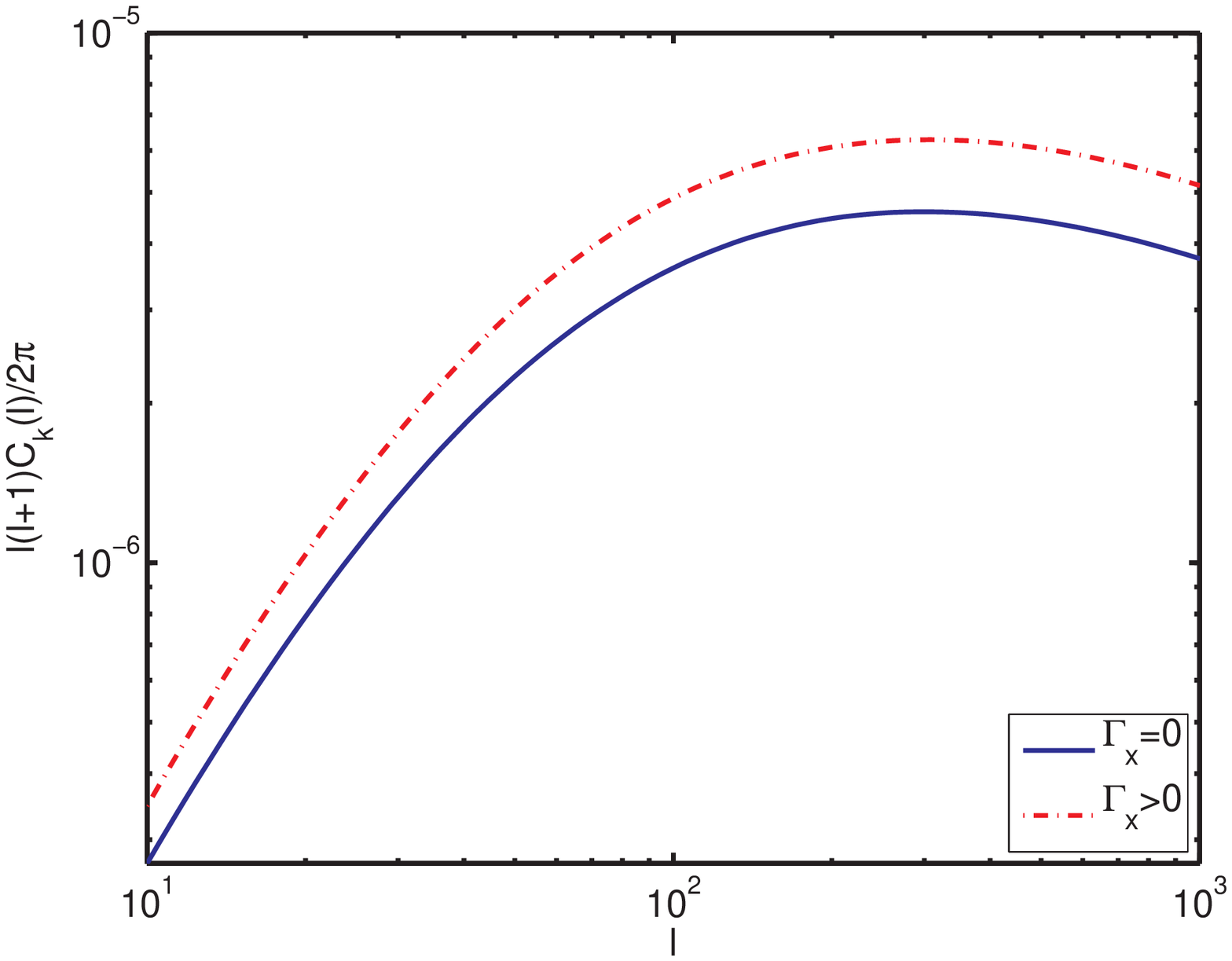}\hfill
\includegraphics[width=0.50\textwidth]{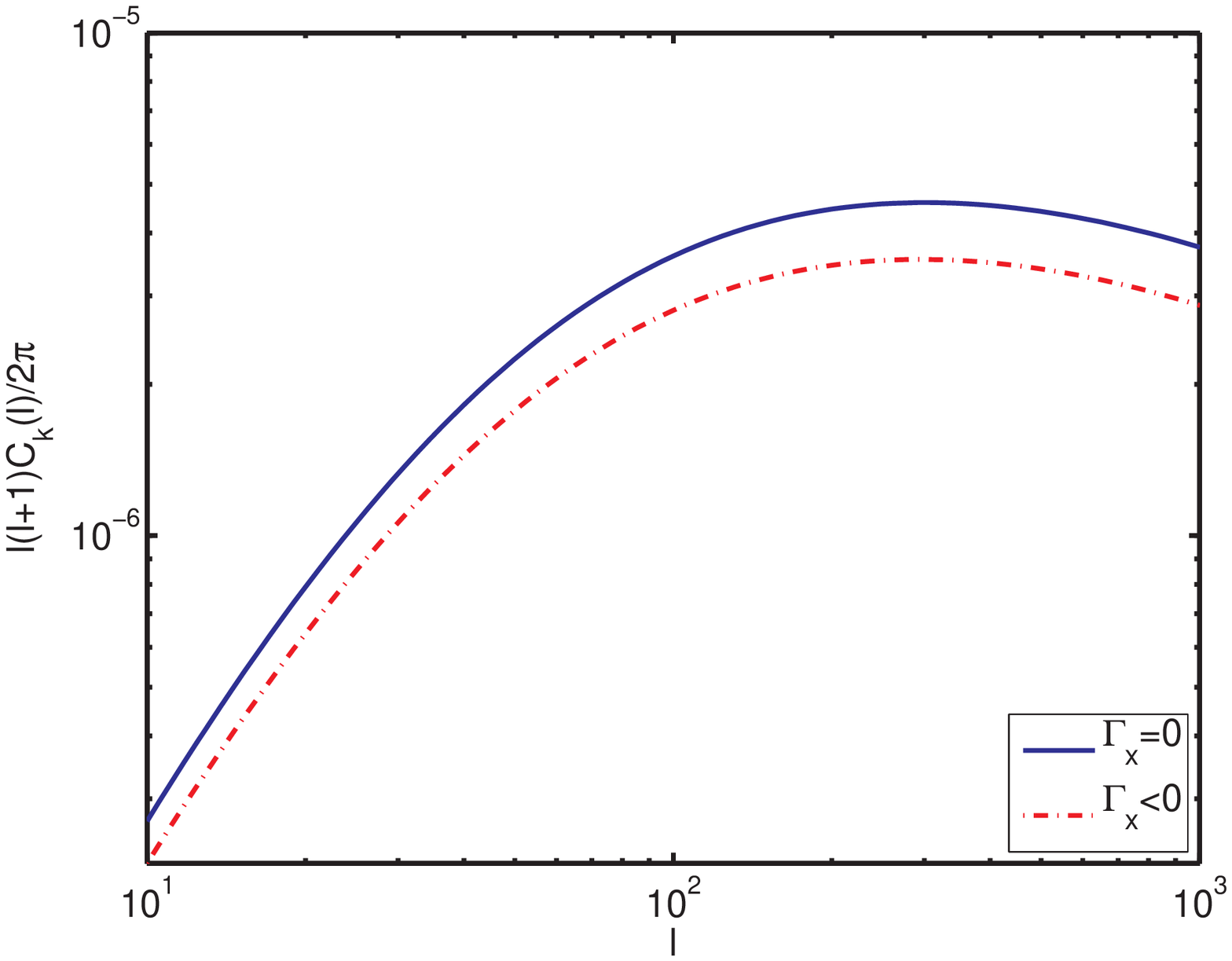}
\caption{Weak lensing convergence power spectra $\ell(\ell+1) C_k(\ell)
/2\pi$ in the interacting models (dashed-dotted lines), with
$\Gamma_x=+ 0.3H_0, -0.2H_0 $, in comparison to non-interacting
models (solid lines). } \label{lensingx}
\end{figure*}

Using the growth function in Eq.~(\ref{lensing}), we compute the
impact of interacting dark energy on the weak lensing convergence power
spectra. The results are shown in Fig.~\ref{lensingx}: given the normalization to today's values of $\Omega_{c0}$ and $\delta_{c0}$, the model
with $\Gamma_x>0$ ($\Gamma_x<0$) exhibits enhanced (suppressed)
power relative to the non-interacting model. The cause of this is both the background effect, i.e., since the matter density in the past was higher (lower)
than the non-interacting case, and the additional reinforcing effects of modified Hubble friction and modified Newton constant. These effects work together to produce
stronger (weaker) gravitational potentials and hence a
stronger (weaker) light deflection.

\section{Conclusions}

For the interacting dark energy models described by the covariant
interaction four-vector
\begin{eqnarray}
Q_c^\mu &=& Q_c u_c^\mu\,, \\ Q_c &=&-\Gamma_c \bar{\rho}_c(1+
\delta_c) -\Gamma_x \bar{\rho}_x(1+ \delta_x),
\end{eqnarray}
we investigated the effect of the interaction on the background
evolution of densities, on the growth factor in structure
formation, and on the weak lensing convergence power spectrum.

We considered the two cases, (I)~$\Gamma_x=0$ and
(II)~$\Gamma_c=0$ separately. In the general case where both transfer rates are nonzero, the results are a linear superposition of the two cases, and will depend on the signs and relative magnitudes of $\Gamma_A$ [the density perturbations will be given by Eq.~(\ref{evolutionxred})].

Case~I does not produce changes in the
evolution equation for the dark matter density perturbations, so
that the change in the growth factor and weak lensing is due
purely to the different background evolution. In case~II, the
$\delta_c$ evolution equation has two additional modifications: a
modified Hubble friction term, and a modified source term,
expressed as a modified effective Newton constant for dark matter:
\begin{equation}
\ddot \delta_c+2H\!\left(\!1-\frac{\Gamma_x}{H}
\frac{\bar\rho_x}{\bar\rho_c}\! \right)\!\dot \delta_c-4\pi G_{\rm
eff}\bar\rho_c\delta_c -4\pi G \bar\rho_b\delta_b=0\,,\label{evdpx2}
\end{equation}
where
\begin{equation}
{G_{\rm eff} \over G} =
1+\frac{2}{3a}\frac{\Gamma_x}{H}\frac{\bar\rho_x}{\bar\rho_c}
\left\{2-3w+\frac{\Gamma_x}{H}\left(1+\frac{\bar\rho_x}{\bar\rho_c}
\right)\right\}.
\end{equation}
These modifications act in the same direction as the modification
to the background evolution.

Both cases present the same qualitative feature in the background evolution and in the growth of structure, assuming a normalization to today's values of $\Omega_{c0}$ and $\delta_{c0}$:
for energy transfer from dark matter to dark energy, i.e. $\Gamma_A>0$, the dark matter density is higher in the past
relative to the non-interacting case, and hence the gravitational
potential is higher, leading to an enhancement of growth and a
stronger lensing signal.
By contrast, when the energy transfer is from dark energy to dark
matter, $\Gamma_A <0$, the dark matter density is
lower in the past, leading to a suppression of growth and a
reduced lensing signal. In the case $\Gamma_c=0$, the effects of the background density evolution are reinforced by the modifications to the Hubble friction and source terms in the density perturbation equation~(\ref{evdpx2}).

Our conclusions are subject to some caveats:
\begin{itemize}
\item
The enhancement or suppression of growth and lensing power relative to the non-interacting case is tied to the assumption that the current values of $\Omega_{c0}$ and $\delta_{c0}$ are equal in the two cases. This gives an accurate picture of how the effects of the interaction operate -- but we cannot conclude that there is relative enhancement or suppression, since in reality the current values of $\Omega_{c0}$ and $\delta_{c0}$ will {\em not} typically be equal. The best-fit values, based on a range of observations, are likely to be different, as shown for other interaction models in~\cite{Lee:2006za,He:2008tn,Bean:2008ac,LaVacca:2008kq,Vergani:2008jv,
LaVacca:2009yp,Gavela:2009cy} and for the $\Gamma_x=0$ model in~\cite{ValiviitaMaartensMajerotto-in-prep}.

\item
We have followed the standard practice of imposing the Newtonian approximation on sub-Hubble scales: for scales close to the Hubble radius, correction terms would start to appear, and numerical integration of the exact perturbation equations~(\ref{da})--(\ref{poisson_complete}) would be necessary.

\item
The weak lensing signal that we have computed is strictly only valid on linear scales. On nonlinear scales, the effects of the interaction would need to be computed (see, e.g.~\cite{Manera:2005ct,Mainini:2006zj,Abdalla:2007rd,Sutter:2008fs,
Mainini:2009cd}) in order to construct an accurate lensing signal, and this could involve N-body simulations for interacting dark matter~\cite{Maccio:2003yk,Baldi:2008ay,Pettorino:2008ez}.

\end{itemize}

Finally, we note that the integrated Sachs-Wolfe effect will also carry an imprint of the dark sector interaction, and this is investigated for the $\Gamma_x=0$ model in~\cite{Schaefer:2008qs,ValiviitaMaartensMajerotto-in-prep}, and for a different interaction model in~\cite{Olivares:2008bx}.

\begin{acknowledgments}
GC-C is supported by the Programme Alban (the European Union
Programme of High Level Scholarships for Latin America),
scholarship No. E06D103604MX, and the Mexican National Council for
Science and Technology, CONACYT, scholarship No. 192680. The work
of RM was supported by the UK's Science \& Technology Facilities
Council. The work of BMS is supported by the German Research Foundation (DFG) within the framework of the excellence initiative through the Heidelberg Graduate School of Fundamental Physics. RM thanks the Cosmology Group at the University of Cape Town, where part of this work was done, supported by a joint Royal Society (UK) and National Research Foundation (South Africa, UID65329) exchange grant on Dark Energy. We thank Kazuya Koyama and Yong-seon Song for helpful discussions. GC-C thanks Jim Cresswell and Kelly Nock for very helpful advice on numerical issues.
\end{acknowledgments}

%\newpage
%\bibliography{ide_dm}

%\bibliographystyle{JHEP}
\bibliography{ide_dm}

\end{document}